\documentstyle[aps,epsf,preprint]{revtex}

\def\[{\left[}
\def\]{\right]}
\def\nn{\nonumber}
\def\({\left(}
\def\){\right)}
\def\labels#1{\label{#1}}
\def\eq#1{(\ref{#1})}

\def\l{\lambda}

\def\.{\cdot}

\def\.{\!\cdot\!}
\def\bi{\begin{itemize}}

\def\ei{\end{itemize}}
\def\be{\begin{eqnarray}}
\def\ee{\end{eqnarray}}
\def\bn{\begin{enumerate}}
\def\en{\end{enumerate}}
\def\h{{1\over 2}}

\def\nn{\nonumber}
\def\l{\lambda}

\def\r2{\sqrt{2}}

\def\x{\times}

\def\eq#1{(\ref{#1})}

\def\t{\tau}

\def\rr2{{1\over\sqrt{2}}}
\def\k{\beta}
\def\bm{\boldmath}
\def\ubm{\unboldmath}
\def\eV{{\rm eV}}
\def\Z{{\bf Z}}
\def\tn{\tilde\nu}
\def\b#1{\overline{#1}}

\begin{document}

\title{Large Mixing Induced by the Strong Coupling with a Single Bulk Neutrino}
\author{C.S. Lam}
\address{Department of Physics, McGill University\\
3600 University St., Montreal, Q.C., Canada H3A 2T8\\
Email: Lam@physics.mcgill.ca}
\maketitle

\begin{abstract}Neutrino is a good probe of extra dimensions. 
Large mixing and the apparent
lack of very complicated oscillation patterns may be an indication of large
couplings between the brane and a single bulk neutrino.
A simple and realistic five-dimensional model of this kind is discussed.
It requires a sterile in addition to three active 
neutrinos on the brane, all coupled strongly
to one common bulk neutrino, but not directly
among themselves.  Mindful that
sterile neutrinos are disfavored in the atmospheric and solar data,
we demand induced mixing to occur 
among the active neutrinos, but not between the active and the sterile.
The size $R$ of the extra dimension is arbitrary in this model, otherwise
it contains six parameters which can be used to fit the three neutrino
masses and the three mixing angles. However, in the model
those six parameters must be
suitably ordered, so a successful fit is not guaranteed. It turns out that
not only the data can be fitted, but as a result of the ordering, a natural
connection between the smallness of the reactor angle $\theta_{13}$ and
the smallness of the mass-gap ratio 
$\Delta M^2_{solar}/\Delta M^2_{atmospheric}$ can be derived.

\end{abstract}

\section{Introduction}
Neutrinos are different from the other fermions. They have small masses
and large mixing. 
Small masses can be explained by the seasaw mechanism,
or by the presence of large extra dimensions \cite{ADDMDDG}.
Large mixing is usually obtained with a right texture, which
in the presence of extra dimensions can be arranged through
 appropriate couplings
between the three or four brane neutrinos and the three or four 
bulk neutrinos. In this paper, we discuss a new and economical
 mechanism to generate large mixing \cite{LN}, through the strong coupling of
a single bulk neutrino to the brane neutrinos.

We take the point of view that 
the main difference between neutrinos and other
Standard Model (SM) fermions is due to  the presence
of extra dimensions. Being SM singlets, sterile neutrinos can roam in
the bulk whereas other fermions cannot.
If SM quantum numbers are trapped on the
brane, perhaps flavor and generation
numbers are trapped there as well, 
in which case there is no reason
to require more than one neutrino in the bulk. With only one bulk
neutrino to couple to, textures can no longer be arranged in the usual way,
so the question  is
whether large and correct mixing between the brane neutrinos can be 
automatically induced
by their couplings with this one common bulk neutrino.
The answer is clearly no if the brane-bulk couplings are weak, 
but there is a chance when they are strong.

To be definite we consider a simple model with 
one extra flat dimension of radius $R$, which we use as the
inverse-energy unit in the problem.
This model contains $f$ brane neutrinos,
each with its own Majorana mass $m_i/R$, and a single sterile bulk neutrino,
coupled via Dirac masses $d_i/R$ to the brane neutrinos.
The brane neutrinos do not directly couple among themselves, and the bulk
neutrino does not have a 5D mass.
All phases are ignored
in this simple model, and for technical simplicity
we shall take all $m_i$ to be positive, and not equal to an integer
nor a half integer. The 
$f=3$ case of this model with equal couplings $d_i$ 
was first proposed in Ref.~\cite{DS}.

We will not ask in this paper the dynamical origin of the Majorana mass nor the
coupling; we will simply take the $2f$ dimensionless parameters 
$m_i$ and $d_i$ to be real and adjustable. Unless that works 
phenomenologically, there is no point to speculate on the dynamical
origin of the model.

Models of this kind have been studied by many authors 
\cite{M1,M2,M3}, but often
without the Majorana masses, with $f=1$, and/or in the
weak coupling limit. When a single brane neutrino couples weakly
to the bulk, oscillations occur between the brane neutrino and only a few
low-lying Kaluza-Klein (KK) modes of the sterile bulk. 
Nevertheless, the participation
of the few excited KK modes is important as they improve the poor
energy dependence for a solar $\nu_e$ to oscillate 
into a sterile neutrino \cite{M1}.
To explain also atmospheric oscillation, it is necessary to introduce
additional bulk neutrinos and arrange the texture of 
the couplings appropriately.

When couplings are increased, more KK modes are
involved, which causes a wild and complicated oscillation pattern
to emerge \cite{M2}.
This could have been a smoking gun for the presence
of an extra dimension, but
unfortunately there is no hint in the experimental data to support it. 
The probability to convert an active neutrino into a sterile KK
mode is also enhanced, a fact which might cause some discomfort because
oscillation into sterile neutrinos is disfavored by 
experiments\footnote{The LSND experiment will be ignored 
throughout this paper.}.
In the strong coupling limit when $f=1$, the worrisome wild oscillations
disappear, because the widths of the states are now larger than their
separations, making the spectrum effectively a continuum.
Leakage into the KK modes is now total, 
with a leakage rate increasing with the coupling \cite{M3},
so very quickly the single brane neutrino simply disappear in sight.
If
$f>1$, leakage is still present but is now divided among the $f$ brane 
neutrinos. In addition to the continuum in the spectrum, 
there are now $f-1$ isolated
eigenstates through which brane neutrinos with induced mixing
can oscillate. For $f=3$, there are 2 isolated eigenvalues
and hence only one mass difference, making it impossible to explain
solar and atmospheric oscillations simultaneously \cite{LN}.

In this paper we show that the strong coupling limit of $f=4$ works.
Besides the three active neutrinos $\nu_e,\nu_\nu,\nu_\t$
on the brane, we must introduce
a fourth one, $\nu_s$, which is sterile.
Bearing in mind that oscillation into sterile neutrinos is disfavored,
we concentrate on a scenario where no induced mixing exists between
the sterile and the active brane neutrinos. 
In that case 
it turns out that the leakage into
the bulk comes solely from $\nu_s$,
which is nice for two reasons. First, since $\nu_s$ is sterile,
there seems to be no reason for it to be confined to the brane, so 
one might expect it to mix more easily with
the bulk. Second and more importantly,
without leakage into the bulk or mixing with $\nu_s$, the active
neutrinos behave as if the fifth dimension were absent.
That increases the chance for the model to work because
we know that 4D analyses are phenomenologically successful. 
Nevertheless, we must realize that
the presence of the fifth dimension  is indispensable for 
generating the induced mixing between the active neutrinos.

We will show that this decoupling scenario can be obtained  by choosing
$m_4$, the largest of the Majorana masses, to be much larger than the other
$2f-2=6$ dimensionless parameters. 
Six is just the right number of parameters
to fit the three mixing angles and the three dimensionless
neutrino masses $M_\alpha R$, whatever the radius $R$ is. 
Nevertheless, there is
no guarantee that we can get a fit, because the resulting parameters 
must be ordered in a special way to be discussed later.
It is therefore encouraging that a
fit to experimental data is actually possible, with the requisite
ordering of the parameters maintained.

As a consequence of this ordering, we will show that a
very interesting connection
can be established between the smallness of
the reactor angle $\theta_{13}$ and the smallness of the
mass-gap ratio $\Delta M^2_{solar}/\Delta M^2_{atmospheric}$.

After reviewing the known result of the strong coupling limit \cite{LN} 
in Sec.~2 for any $f$,
a new parameterization is introduced in Sec.~3, which is needed for the
phenomenology application in Sec 4.

\section{The Model and Its Strong Coupling Limit}
The model to be considered here consist of $f$ brane
neutrinos, with Majorana masses $m_i/R$, and a single massless bulk neutrino
which is coupled to the brane neutrinos through Dirac masses $d_i/R$, 
with $R$ being the
radius of the extra dimension. In units of $1/R$, its mass matrix is 
given by\cite{DS,LN}
\be M=\pmatrix{m_1&0&\cdots&0&d_1&d_1&d_1&d_1&d_1&\cdots\cr
               0&m_2&\cdots&0&d_2&d_2&d_2&d_2&d_2&\cdots\cr
               \cdots&&\cdots&&&&&&&\cdots\cr
                 0&0&\cdots&m_f&d_f&d_f&d_f&d_f&d_f&\cdots\cr
               d_1&d_2&\cdots&d_f&0 &0 &0 &0 &0 &\cdots\cr
               d_1&d_2&\cdots&d_f&0 &1 &0 &0 &0 &\cdots\cr
               d_1&d_2&\cdots&d_f&0 &0 &-1 &0 &0 &\cdots\cr
               d_1&d_2&\cdots&d_f&0 &0 &0 &2 &0 &\cdots\cr
               d_1&d_2&\cdots&d_f&0 &0 &0 &0 &-2 &\cdots\cr
               \cdots&&\cdots&&&&&&\cdots&\cdots\cr},\nn\\ \labels{mm}\ee
in which the rows and columns are labelled respectively by the left-handed
and right-handed neutrinos.
The first $f$ rows and columns are the brane neutrinos, and the rest
are the $n$th mode of the bulk neutrino. The $2f$ parameters $d_i$ and
$m_i$ are real and arbitrary, but for technical simplicity 
we will assume the $m_i$'s
to be non-negative, and not equal to an integer or a half-integer.

The eigenvalues $\l$ of this matrix satisfies the characteristic equation
\be
{1\over\pi}\tan(\pi\l)=d^2\,r(\l),\labels{char}\ee
where $d^2=\sum_{i=1}^fd_i^2$, so $e_i=d_i/d$ obeys the constraint
\be
\sum_{i=1}^fe_i^2=1.\labels{ei}\ee
The function $r(\l)$ is given by
\be
r(\l)=\sum_{i=1}^f{e_i^2\over\l-m_i}.\labels{r}\ee

In the absence of coupling, the bulk eigenvalues are 
integers and the brane eigenvalues are located 
at $\l=m_i$. Accordingly we label the flavor states by
an index $i$ or $n$, where $i=1,\cdots,f$ label the brane states
and $n\in\Z$ label the KK tower of bulk states.

In the strong-coupling limit\footnote{
Strong coupling is expected if 
$R^{-1}$ is the only additional energy scale that is relevant in the problem.
This may be the case if there is only one large extra dimension, because then
the gravitational energy scale is still very high and presumably irrelevant.
Neutrino mass is
given by the formula  $M=\kappa v/\sqrt{R}$ \cite{ADDMDDG}, 
where $v$ is the Higgs 
expectation value and $\kappa$ is the brane-bulk coupling constant
in the Lagrangian. $\kappa$ has a dimension of $\sqrt{R}$, and therefore can
be expected to be of that order if there is no other relevant
dimensional quantity around in the bulk. 
Our coupling constant $d_i=MR=\kappa v\sqrt{R}\sim
vR$ is then much larger than 1, if $R$ is of a sub-milliliter size.}, 
when $d\gg m_i$ and 1,
the characteristic equation \eq{char} is satisfied
when either $\tan(\pi\l)=\infty$, or $r(\l)=0$. 
The former produces half-integer eigenvalues $\l_n=n+\h,\ n\in\Z$. The
latter produces $f-1$ `isolated' eigenvalues
$\l_\alpha\ (1\le \alpha\le f-1)$, 
which are the roots of that equation\footnote{There is 
a slight change of notation from Ref.~\cite{LN}.
$\l_2,\cdots,\l_f$ in Ref.~\cite{LN} have been
renamed $\l_1,\cdots,\l_{f-1}$. Accordingly the first column of 
$V$ in eq.~(29)
of Ref.~\cite{LN} now appears at the last column.}.
Since $r(\l)$ approaches $+\infty$ for $\l=m_i+$ and $-\infty$
for $\l=m_i-$, it
is not difficult to see that the isolated eigenvalues must always
occur between pairs of $m_i$'s. In other words, 
\be
0<m_1<\l_1<m_2<\l_2<\cdots<\l_{f-1}<m_f.\labels{order}\ee

It is important to realize that in the strong coupling limit,
flavor states are so thoroughly mixed up that it is impossible to
tell whether a mass eigenstate is more brane like or bulk like, much
less which flavor brane neutrino it should be identified with. In particular,
it is not true that the states with half-integer eigenvalues are bulk like
and the isolated ones are brane like. After all, there are only $f-1$
isolated eigenvalues and not $f$.

The transition amplitude ${\cal A}_{ij}(\t)$ 
from a brane flavor neutrino $j$ of energy $E$
to a brane flavor neutrino $i$, 
after it has traversed a distance $L=2E\t R^2$, 
is given by
\be{\cal A}_{ij}(\t)&=&\sum_{\l}u_{i\l}u^*_{j\l}e^{-i\l^2\t},\labels{aij}\ee
where the infinite-dimensional
mixing matrix $u_{i\l}$ is given by the flavor-$i$ component
of the normalized mass eigenstate with eigenvalue $\l$.
The sum over $\l_n=n+\h$ can be computed in the strong-coupling limit, 
resulting
in a term proportional to a function $g(x)$, with $g(0)=1$
and $g(x)\to (1-i)/\sqrt{2\pi x}$ for large $|x|$. The final result 
is \cite{LN},
\be{\cal A}_{ij}(\t)&=&
\sum_{\alpha=1}^{f-1}V^*_{i\alpha}V_{j\alpha}
e^{-i\l_\alpha^2\tau}+V^*_{if}V_{jf}g(K^2\t),
\labels{ampfin}\ee
where $K^2=d^2(1+d^2\pi^2)\gg 1$,
$V_{i\alpha}=u_{i\l_\alpha}$, and $V_{if}=e_i^2$. Explicitly,
the $f\x f$ induced mixing matrix is given by
\be
V&=&\pmatrix{e_1x_{11}/\sqrt{s_1}&e_1x_{12}/\sqrt{s_2}&\cdots& 
e_1x_{1,f-1}/\sqrt{s_{f-1}}&e_1\cr
e_2x_{21}/\sqrt{s_1}&e_2x_{22}/\sqrt{s_2}&\cdots& 
e_2x_{2,{f-1}}/\sqrt{s_{f-1}}&e_2\cr
\cdots&&&&\cdots\cr
e_fx_{f1}/\sqrt{s_1}&e_fx_{f2}/\sqrt{s_2}&\cdots& 
e_fx_{f,f-1}/\sqrt{s_{f-1}}&e_f\cr},\labels{v}\ee
where $x_{i\alpha}=1/(\l_\alpha-m_i)$ and
\be
s_\alpha=\sum_{i=1}^f {e_i^2\over(\l_\alpha-m_i)^2}.\labels{s}\ee

The matrix $V$ can be shown to be unitary; actually real orthogonal
because $e_i$ and $m_i$ are assumed to be real.
If we replace the function $g(K^2\t)$ in \eq{ampfin} by 
$\exp(-i\l_f^2\t)$, for some real number $\l_f$, then 
\eq{ampfin} 
is the usual 4D transition-amplitude formula
from flavor neutrino $n_j$ to $n_i$, through the mass eigenstates
$\tilde n_k$, with
a mixing matrix\footnote{This relation is usually written as
$\nu_i=\sum_{j=1}^3U_{ij}\tilde\nu_j\ (i=e,\mu,\tau)$ for three flavors.
We use this unconventional notation to avoid later confusion,
because $n_i$ turns out to be
a permutation of $\nu_i$, and $\tilde n_j$ a permutation of $\tilde n_j$,
so $U$ is obtained from $V$ by permutating rows and columns.}
 $V$: $n_j=\sum_kV_{jk}\tilde n_k$. Moreover, 
if we choose $\l_f\gg \l_\alpha$, for $1\le\alpha\le f-1$, then 
for any detector with finite resolution, the 
last term $\exp(-i\l_f^2\t)$ averages to zero except for tiny
$\t$'s, just like $g(K^2\t)$. In other words, 
this 5D model in the strong coupling limit
resembles a 4D theory with a mixing matrix $V$ and a mass
pattern $(f-1)+1$, with a huge mass gap between the first
$f-1$ neutrinos $\tilde n_\alpha$ and the last one, $\tilde n_f$.
Since $\tilde n_f$ represents the total effect of the
tower of half-integral eigenvalues, rather than 
oscillating through it, 
it will be shown immediately below that a brane neutrino disappears into it.
In other words,  
we should probably think of $\l_f^2$ as having a negative imaginary part. 
 
Unlike the 4D case, the total probability for an initial
flux of flavor $j$ to remain on the
brane is not 1. There is a leakage into the bulk, equal to
\be
1-\sum_{i=1}^f|{\cal A}_{ij}(\t)|^2=
e_j^2\{1-|g(K^2\t)|^2\}\simeq e_j^2.\labels{leak}\ee
The leakage starts out to be 0 at $\t=0$, but in the strong-coupling
limit it reaches its asymptotic value $e_j^2$ 
almost instantaneously. This leakage is given by the square of the
matrix elements in the last column of $V$. Since $\sum_je_j^2=1$,
the total leakage into the bulk from all brane neutrinos is 1, independent
of $f$.

\section{A Simple Parameterization}
It is fairly complicated to compute the elements of $V$ in \eq{v}.
Given the parameters 
$e_i$ and $m_i$, we must first find the eigenvalues $\l_\alpha$
by solving $r(\l_\alpha)=0$. Then we have to compute 
$x_{i\alpha}$ and $s_\alpha$, and finally the matrix elements of $V$.
In this section, we discuss a much simpler parameterization of
$V$ which enables its elements to
 be computed without having to solve any equation.
This new parameterization is needed for the phenomenological
discussion in the next section.

Instead of the $2f-1$ parameters $e_i$ and $m_i$, we use as independent
parameters $m_i$ and $\l_\alpha$. 
Since $\l_\alpha$ are the zeros of the function $r(\l)$ in \eq{char}, 
they must obey the ordering in \eq{order},
but otherwise these parameters are free to vary.
To show that we must be able to compute $e_i^2$ from these new parameters.
To that end note that the function $r(\l)$
is a meromorphic function with zeros at $\l_\alpha$ and simple poles at $m_i$,
and it goes like $1/\l$ for large $|\l|$.
Hence it is equal to
\be
r(\l)={\prod_\k(\l-\l_\k)\over\prod_{j}(\l-m_j)}.
\labels{ralt}\ee
Unless otherwise specified, all products over $j$ are taken from 
1 to $f$, and all products over $\k$ are taken from 1 to $f-1$.
The quantity $e_i^2$, being the residue at the simple pole
$m_i$, is then equal to
\be
e_i^2={\prod_{\k}(m_i-\l_\k)\over\prod_{j\not=i}(m_i-m_j)}.
\labels{ei2}\ee
With  the ordering \eq{order},
$e_i^2$ obtained from \eq{ei2} are automatically non-negative.

We can now calculate from \eq{s} to get
\be
s_\alpha=(-1)^\alpha{\prod_{\k\not=\alpha}(\l_\alpha-\l_\k)\over 
\prod_j(\l_\alpha-m_j)},
\labels{salt}\ee
which is always positive if \eq{order} is obeyed.
Since $x_{i\alpha}=1/(\l_\alpha-m_i)$, the matrix elements of $V$ become
\be
V_{i\alpha}&=&{1
\over\l_\alpha-m_i}
\sqrt{\Bigg|{\prod_\k(m_i-\l_\k)\prod_j(\l_\alpha-m_j)\over
\prod_{j\not=i}(m_i-m_j)\prod_{\k\not=\alpha}(\l_\alpha-\l_\k)}\Bigg|}\nn\\
V_{if}&=&e_i^2={\prod_{\k}(m_i-\l_\k)\over\prod_{j\not=i}(m_i-m_j)}.\labels{valt}
\ee
The index $i$ runs from 1 to $f$, and
the index $\alpha$ runs from 1 to $f-1$.
It is easy to see that the sign of $V_{ij}$ is positive unless $j<i$, when
 it is negative. With the sign taken care of,
the magnitude of the matrix elements can be written in the more symmetric form
\be
\big|V_{i\alpha}\big|&=&
\sqrt{\Bigg|{\prod_{\k\not=\alpha}(m_i-\l_\k)\prod_{j\not=i}(\l_\alpha-m_j)\over
\prod_{j\not=i}(m_i-m_j)\prod_{\k\not=\alpha}(\l_\alpha-\l_\k)}\Bigg|}\nn\\
V_{if}&=&e_i^2={\prod_{\k}(m_i-\l_\k)\over\prod_{j\not=i}(m_i-m_j)}.
\labels{valt1}\ee
Note that $V$ is scale invariant, in that
$V$ is unchanged 
if we multiply every parameter $m_i,\l_\alpha$ by the same scale factor.

The matrix elements of $V$ in \eq{valt1} appear deceptively symmetrical,
but they are actually not because of \eq{order}. 
For example, there is no way to let
$m_3-m_2\to 0$ without also letting $\l_2-m_2$ and $m_3-\l_2\to 0$. Similarly,
there is no way to make $m_3$ large without also 
making $\l_3$ and $m_4$ large, but we can make $m_4$ large without affecting
other parameters.
 
In view of \eq{order}, all quantities can be expressed in terms
of the $2(f-1)$ consecutive distances 
\be
p_\alpha&=&\l_{\alpha}-m_\alpha\ ,\qquad (1\le \alpha\le f-1)\nn\\
q_\alpha&=&m_{\alpha+1}-\l_{\alpha}\ .\labels{dist}\ee
Since $V_{ij}$ is scale invariant, there are actually only
$2f-3$ independent parameters needed to specify $V$.

From \eq{valt1} one sees that every matrix element of $V$ has an
equal number of $m_f$'s in the numerator as the denominator, except
$V_{\alpha f}$ and $V_{f\alpha}$, which have one more $m_f$ in the
denominator than the numerator. Hence if $m_f\gg m_\alpha, \l_\alpha$,
the induced mixing matrix $V$ will take on the form
\be
\pmatrix{
*&*&\cdots&*&0\cr
*&*&\cdots&*&0\cr
\cdots&\cdots&\cdots&\cdots&\cdots\cr
*&*&\cdots&*&0\cr
0&0&\cdots&0&1\cr},\labels{vasterisk}\ee
where the asterisks indicate elements that are generally not zero. 
In that case, there
is no induced mixing between the first $f-1$ flavor neutrinos and the
last. There is also no leakage of $n_\alpha$ into the bulk, for
$1\le\alpha\le f-1$.

\section{Phenomenology}
The neutrino mass corresponding to  eigenvalue $\l_\alpha$ 
is $M_\alpha=\l_\alpha/R$.
In order to have two different $\Delta M^2$'s to explain both atmospheric
and solar neutrino oscillations, we need three distinct $\l_\alpha$'s,
hence $f=4$.
Thus in additional to the three brane neutrinos $\nu_e,\nu_\mu,\nu_\t$
which are active,  we are forced to have one more brane neutrino $\nu_s$
that is sterile. 

In the usual 4D analysis, a sterile neutrino $\nu_s$  is introduced to get us
three independent $\Delta M^2$'s, needed to explain solar,
atmospheric, as well as LSND oscillations. In the
present context in five dimensions, leakage into the bulk is simulated by
an infinite $\Delta M^2$, so we need four brane
neutrinos just to have two finite $\Delta M^2$'s.

Experimentally, oscillation from solar and atmospheric neutrinos into 
the sterile ones  are disfavored. Since $\nu_s$ must be present in the
theory, this requirement can be fulfilled if $\nu_s$
decouples from the active neutrinos, as in \eq{vasterisk}.  
All we have to do is to let $n_4=\nu_s$,
and choose $m_4$
to be far greater than $\l_\alpha$ and $m_\alpha$.
 
Since $V$ is scale invariant, it is more useful to express 
it in scale-invariant parameters
\be
\xi&=&p_1/(q_1+p_2+q_2+p_3)\nn\\
\alpha&=&q_1/(q_1+p_2+q_2+p_3)\nn\\
\beta&=&p_2/(q_1+p_2+q_2+p_3)\nn\\
\gamma&=&q_2/(q_1+p_2+q_2+p_3)\nn\\
\delta&=&p_3/(q_1+p_2+q_2+p_3).\labels{abcd}\ee
Only four of these are independent because
\be
\alpha+\beta+\gamma+\delta=1.\labels{sum1}\ee
See Fig.~1 for a summary of the parameters used.

The exact expressions for the asterisks in \eq{vasterisk} are:
\be
V_{11}&=&
\sqrt{(\alpha+\xi+\beta)(1+\xi)(\gamma+\alpha+\beta)\alpha
\over(\alpha+\xi)(\gamma+\alpha+\xi+\beta)(\beta+\alpha)}
\nn\\
V_{12}&=&
\sqrt{\xi(1+\xi)\gamma\beta
\over(\alpha+\xi)(\gamma+\alpha+\xi+\beta)(\delta+\gamma)(\beta+\alpha)}\nn\\
V_{13}&=&
\sqrt{\xi(\alpha+\xi+\beta)\delta(\gamma+\beta+\delta)
\over(\alpha+\xi)(\gamma+\alpha+\xi+\beta)(\delta+\gamma)}
\nn\\
V_{21}&=&
-\sqrt{\beta(\gamma+\beta+\delta)(\gamma+\alpha+\beta)\xi
\over(\alpha+\xi)(\gamma+\beta)(\beta+\alpha)}\nn\\
V_{22}&=&
\sqrt{\alpha(\gamma+\beta+\delta)\gamma(\alpha+\xi+\beta)
\over(\alpha+\xi)(\gamma+\beta)(\delta+\gamma)(\beta+\alpha)}\nn\\
V_{23}&=&
\sqrt{\alpha\beta\delta(1+\xi)
\over (\alpha+\xi)(\gamma+\beta)(\delta+\gamma)}\nn\\
V_{31}&=&
-\sqrt{\gamma\delta\alpha\xi
\over(\gamma+\alpha+\xi+\beta)(\gamma+\beta)(\beta+\alpha)}
\nn\\
V_{32}&=&
-\sqrt{(\gamma+\alpha+\beta)\delta\beta(\alpha+\xi+\beta)
\over(\gamma+\alpha+\xi+\beta)(\gamma+\beta)(\delta+\gamma)(\beta+\alpha)}
\nn\\
V_{33}&=&
\sqrt{(\gamma+\alpha+\beta)\gamma(\gamma+\beta+\delta)(1+\xi)
\over(\gamma+\alpha+\xi+\beta)(\gamma+\beta)(\delta+\gamma)}\nn\\
\labels{vij}\ee
We shall henceforth denote this $3\x3$ sub-matrix of $V$ as $\b V$.

With phases neglected, there are six measurable quantities, 
the three neutrino masses
 and the three mixing angles.
There are seven parameters in the 
model, so one of these parameters can never be determined. We shall take it
to be the radius $R$ of the fifth dimension. 
The other six parameters $m_\alpha,
\l_\alpha$ can in principle
 be determined from the six experimental quantities, but 
there is no guarantee that the fitted parameters
will obey the ordering relation \eq{order}. It is therefore very 
encouraging that the resulting fit from the known experimental
data does obey \eq{order}.

As seen in \eq{vij}, four parameters are needed to specify the induced
mixing matrix $\b V$, not three. Hence the values of the neutrino masses
$M_\alpha$ may affect the values of the mixing angles. This turns out
to be the case in a particularly relevant 
situation which  we shall discuss later.

The fitting of experimental data is not as straight-forward as it might be,
because we do not know how $V$ is linked to the experimental mixing matrix $U$,
defined by $\nu_i=\sum_{j=1}^3U_{ij}\tilde\nu_j\ (i=e,\mu,\t)$, and
parameterized as
\be
U=\pmatrix{
c_{12}c_{13}& s_{12}c_{13}& s_{13}\cr
-s_{12}c_{23}-c_{12}s_{23}s_{13}&  c_{12}c_{23}-s_{12}s_{23}s_{13}& s_{23}c_{13}\cr
s_{12}s_{23}-c_{12}c_{23}s_{13}& -c_{12}s_{23}-
s_{12}c_{23}s_{13}&c_{23}c_{13}\cr}.\labels{angles}\ee 
As usual, 
$s_{ij}=\sin\theta_{ij}$ and $c_{ij}=\cos\theta_{ij}$, and phases are
ignored. We shall follow the usual convention
 to take all mixing angles to be between 0 and $\pi/2$, so that
both $s_{ij}$ and $c_{ij}$ are non-negative.

We do know that the three $n$'s are flavor neutrinos so they
must be the same as the three $\nu$'s, and that the three $\tilde n$'s
are mass eigenstates so they must be identical to the three $\tilde\nu$'s.
But, we do not know which is which. $n_i$ is arranged according to
increasing Majorana masses $m_i$ of the uncoupled flavor neutrinos, and
$\tilde n_j$ is arranged according to increasing eigenvalues. On the other
hand, $\nu_i$ is fixed by their weak interactions, and $\tilde\nu_j$
is tied to $\nu_i$ through $U_{ij}$ and the experimental mixing angles.
To find the correct assignments of $n_i$ to $\nu_i$ and $\tilde n_j$
to $\tn_j$, we need to determine what permutations of rows and columns
of $\b V$ are needed to bring it into the form of $U$. 
As noted earlier,
although there are six parameters in $\b V$, the ordering requirement
\eq{order} does make the columns and rows of $\b V$ inequivalent.

The present experimental situation is as follows.

Atmospheric mixing is maximal. It gives $\tan^2\theta_{23}=1$,
with a square mass difference 
$\Delta M^2_{atm}=2.5\x10^{-3}\ ({\rm eV})^2$ \cite{SKATM}.

There are several possible solar-neutrino solutions 
\cite{SOLDAT,SOLAR1,SOLAR2}. 
Of these we  consider the three most likely ones:  
LMA, LOW, and VAC. 
The exact parameters and 
the goodness of the fit of each depend on the specific analysis, but generally
speaking, the LMA solution is the most favorable.
In global analyses
where both rate and spectrum are taken into account, the SMA solution is 
poor so we  will not consider it. The parameters do not
vary that much from analysis to analysis, so for the following discussions,
we will adopt the parameters taken from Table 2 of
Ref.~\cite{SOLAR1}. The $(\tan^2\theta_{12},\Delta M^2_{sol})$ values
for these solutions are:
LMA=(0.36, $5\x10^{-5}$), VAC=(0.363, $1.4\x10^{-10}$),
LOW=(0.69, $1.1\x10^{-7}$). The unit for
$\Delta M^2$ is (eV)$^2$.

As for the reactor angle, only an upper
bound of $\tan^2\theta_{13}< 0.04$ is known\footnote{
CHOOZ quotes an upper bound of $\sin^2(2\theta)=0.1$ for large
$\Delta M^2$, corresponding to a $\tan^2\theta=0.026$.  
In the vicinity of $\Delta M^2=2.5\x10^{-3}\ {\rm eV}^2$, the best-fitted
atmospheric mass difference, the bound is $\tan^2\theta\simeq 0.038$. 
We will
be cautious and take $\tan^2\theta_{13}<0.04$.}
from the CHOOZ experiment \cite{CHOOZ}. 

We do not know the values of $M_\alpha$, 
except for an upper  bound of about
2 eV  \cite{MAINZ} from end point measurements of tritium $\beta$-decay.
We will therefore treat $M_1$ as a parameter, to be varied from 0
to 2 eV. Using \eq{order},
the values of $M_2$ and $M_3$ are then determined from the magnitude of the
solar and atmospheric gaps.
If the solar gap is on top, then
$M_2=\sqrt{M_1^2+
(\Delta M^2)_{atm}}$ and $M_3=\sqrt{M_2^2+
(\Delta M^2)_{sol}}$. We will designate this case as $G=1$.
If the solar gap is at the bottom, then
 $M_2=\sqrt{M_1^2+
(\Delta M^2)_{sol}}$ and $M_3=\sqrt{M_2^2+
(\Delta M^2)_{atm}}$. We will designate this case as $G=2$.

Let 
\be
\rho=(M_2-M_1)/(M_3-M_1)=\alpha+\beta=1-\gamma-\delta\labels{ratio}\ee
be the ratio of mass differences. 
Its value depends on $M_1$, on $G$, and on the
 solar-neutrino solution. Listed below are
 some values of $\rho$
for the solar solutions
[LMA, LOW, VAC] at various $M_1$ and $G$:
\be
&&[.98995,\  .99998,\ 1.00000]\quad (M_1=0\,\eV,\ G=1)\nn\\
&&[.98001,\  .99996,\ 1.00000]\quad (M_1=1\,\eV,\ G=1)\nn\\
&&[.98000,\  .99996,\ 1.00000]\quad (M_1=2\,\eV,\ G=1)\nn\\
&&[.14142,\  .00663,\ 0.00024]\quad (M_1=0\,\eV,\ G=2)\nn\\
&&[.02001,\  .00004,\ 0.00000]\quad (M_1=1\,\eV,\ G=2)\nn\\
&&[.02000,\  .00005,\ 0.00000]\quad (M_1=2\,\eV,\ G=2).\labels{rvalue}\ee
We see that $\rho$ is close to 1 for $G=1$,  and
close to 0 for $G=2$.

An analytical solution exists if we
approximate $\rho$ to be 0 for $G=2$, and 1 for 
$G=1$. This is discussed below, separately for $G=2$ and $G=1$, together
with the numerical solutions when this approximation is not made.

\subsection{\bm $G=2$\ubm}
The mass of $\tilde n_i$ increases with $i$.
When the solar gap is at the bottom, either 
$(\tilde n_1,\tilde n_2,\tilde n_3)=(\tn_1,\tn_2,\tn_3)$, or 
$(\tilde n_1,\tilde n_2,\tilde n_3)=(\tn_2,\tn_1,\tn_3)$.

When $\rho=0$, it follows from \eq{ratio} that $\alpha=\beta=0$,
because both $\alpha$ and $\beta$ must be non-negative. Using
\eq{sum1}, we also get $\delta=1-\gamma$. 
Defining $\alpha/\beta=a$, the matrix
$\b V$ in this limit simplifies to
\be
\b V=\pmatrix{
\sqrt{(1+\xi)\gamma a/(\gamma+\xi)(1+a)}&
\sqrt{(1+\xi)\gamma/(\gamma+\xi)(1+a)}&
\sqrt{\xi(1-\gamma)/(\gamma+\xi)}\cr
-\sqrt{1/( 1+a)}& \sqrt{a/(1+a)}& 0\cr
-\sqrt{(1-\gamma)a\xi/(\gamma+\xi)(1+a)}&
-\sqrt{(1-\gamma)\xi/(\gamma+\xi)(1+a)}&
\sqrt{\gamma(1+\xi)/(\gamma+\xi)}\cr}\nn\\
\labels{vg2}\ee

The outstanding feature of this matrix is $\b V_{23}=0$.
Since $\tilde n_3=\tn_3$ and the atmospheric mixing is maximal,
the only way that this could happen is for $n_2=\nu_e$, and
$V_{23}=U_{e3}=s_{13}=0$.

Thus the smallness of $s_{13}$ is naturally
related to the smallness of $\Delta M^2_{sol}/\Delta M^2_{atm}$,
at least
in the sense that we are forced to have $s_{13}=0$ if the mass ratio is zero. 
This remarkable feature of the model is a consequence of the ordering
\eq{order}, which demands both
$\alpha$ and $\beta$ to be non-negative. It is not the result of
an arranged texture.

To implement the atmospheric mixing
condition $\tan^2\theta_{23}=1$, we need $\xi=\gamma/(1-2\gamma)$,
then
\be
\b V=\pmatrix{
\sqrt{a/2(1+a)}& \sqrt{1/2(1+a)}& \sqrt{1/2}\cr
 -\sqrt{1/(1+a)}& \sqrt{a/(1+a)}& 0\cr
 -\sqrt{a/2(1+a)}& -\sqrt{1/2(1+a)}& \sqrt{1/2}\cr}.\labels{vg2a}\ee
To bring $\b V$ into the form of $U$, we can multiply the first column
of $\b V$  by $-1$ (this is just a change of sign convention), 
and change its rows (1,2,3) to (2,1,3). The resulting matrix 
\be
\b U=\pmatrix{
\sqrt{1/(1+a)}& \sqrt{a/(1+a)}& 0\cr
-\sqrt{a/2(1+a)}& \sqrt{1/2(1+a)}& \sqrt{1/2}\cr
  \sqrt{a/2(1+a)}& -\sqrt{1/2(1+a)}& \sqrt{1/2}\cr}\labels{vg2b}\ee
should be compared with $U$ in the approximation $s_{13}=0$, which is
\be
U&=&\pmatrix{
c_{12}& s_{12}& 0\cr
-s_{12}c_{23}&  c_{12}c_{23}& s_{23}\cr
s_{12}s_{23}&-c_{12}s_{23}& c_{23}\cr}.\labels{anglesg2}\ee
It shows $s_{12}=\sqrt{a/(1+a)}$ and 
$c_{12}=\sqrt{1/(1+a)}$.
Since $a$ is arbitrary, all solar solutions can be accommodated.

If we interchange the first two columns and also the last two rows of
$\b U$, we simply interchange 
$a$ with $a^{-1}$. Since $a$ is arbitrary, this form is equally
allowed. Hence there is no way to distinguish the 
case  $(n_2,n_3)=(\nu_\mu,\nu_\t)$ and $(\tilde n_1,\tilde n_2)=(\tn_1,\tn_2)$
from the case $(n_2,n_3)=(\nu_\t,\nu_\mu)$ and 
$(\tilde n_1,\tilde n_2)=(\tn_2,\tn_1)$.

The actual value of $\rho$ is given in \eq{rvalue}. It is zero 
for the VAC solution when $M_1$ is between 1 and 2 eV, to within the
number of digits shown, so the analytical solution discussed above
is perfectly adequate. No more will be said about it.

For the other cases, $\rho>0$, 
numerical solutions are provided in Table I. This Table is constructed
in the following way.
For a given $M_1$ and $\xi$, and a given solar solution, the three parameters
$\beta,\gamma,\delta\ (\alpha=1-\beta-\gamma-\delta)$ are determined
from fitting the three inputs, $\rho$,
$\tan^2\theta_{23}=1$, and $\tan^2\theta_{12}$.
Then $\tan^2\theta_{13}$ is calculated. To test the goodness
of the approximation $\rho=0$, we also list in the last two columns
$a=\alpha/\beta$ and $\gamma/(1-2\gamma)$ calculated from the fitted
parameters. When $\rho=0$,
the former is equal to $\tan^2\theta_{12}$, and the latter is equal
to $\xi$. Other than the LMA solution at $M_1=0$ eV, where
$\rho=.14$ is fairly large, the agreement is quite good. This shows
that the approximation $]rho=0$ is a very reasonable one.

Since $\tan^2\theta_{13}=0$ when $\rho=0$, one might expect 
$\tan^2\theta_{13}$
(and therefore also $\sin^2\theta_{13}$)
to be small when $\rho$ is small. A glance at Table I shows that this is not
necessarily the case, as we can adjust the parameter $\xi$ to
yield a $\tan^2\theta_{13}$ as large as the CHOOZ
bound 0.04, or even bigger. To understand what this means
let us examine the formula
\be
\sin^2\theta_{13}=\b V_{23}^2=
{\alpha\beta\delta(1+\xi)
\over (\alpha+\xi)(\gamma+\beta)(\delta+\gamma)}\labels{whylarge}\ee
obtained from \eq{vij}. For small $\rho$, both $\alpha$ and $\beta$ are small.
As pointed out above, the approximations $\xi\simeq \gamma/(1-2\gamma)$
and $\tan^2\theta_{12}\simeq a=\alpha/\beta$ are good. 
The parameter $a$ is fixed by the solar solution, so $\sin^2\theta_{13}$
is of order $\beta^2$, unless the denominator in \eq{whylarge} is also
small. Now $\delta+\gamma=1-\alpha-\beta\simeq 1$ for small $\beta$
and $\alpha$, and $\xi$ is proportional to $\gamma$ for small $\gamma$,
hence the denominator is small and of order $\beta^2$ if and only if $\gamma$
is itself of order $\beta$. In that case $\xi\simeq\gamma$, and 
\be
\sin^2\theta_{13}\simeq {a\over(a+c)(1+c)}\quad (\gamma=c\beta\ll 1)\labels{
big13}\ee
may indeed be large. An examination of Table I shows this is indeed
what happens: $\tan^2\theta_{13}$ is relatively large only 
when $\gamma$ is relatively small. 

$\gamma$ can vary from 0 to $1-\alpha-\beta\simeq 1$ for small $\rho$.
Except for very small $\gamma$, in most of this range
it yields a small $\tan^2\theta_{13}$. So in a statistical sense 
it is indeed 
true that a small $\rho$ tends to yield a small $\tan^2\theta_{13}$,
meaning that the smallness of the reactor angle
is naturally related to the smallness of the solar to atmospheric
mass gap ratio.

\subsection{\bm $G=1$\ubm}
The main conclusions of the last subsection are essentially unchanged.
Here are the details.

When the solar gap is on top, either
$(\tilde n_1,\tilde n_2,\tilde n_3)=(\tn_3,\tn_2,\tn_1)$ or
$(\tilde n_1,\tilde n_2,\tilde n_3)=(\tn_3,\tn_1,\tn_2)$
must be true.

When $\rho=1$, it follows from \eq{ratio} that $\gamma=\delta=0$
because they must both be positive. In that case it follows from
\eq{sum1} that $\alpha=1-\beta$. Defining $\delta/\gamma=d$, the matrix
$\b V$ in this limit simplifies to be
\be
\b V=\pmatrix{
\sqrt{(1+\xi)(1-\beta)\over1-\beta+\xi}& 
\sqrt{\xi\beta\over(1-\beta+\xi)(1+d)}& 
\sqrt{\xi d\beta\over(1-\beta+\xi)(1+d)}\cr
-\sqrt{\beta\xi\over 1-\beta+\xi}&
\sqrt{(1-\beta)(1+\xi)\over(1-\beta+\xi)(1+d)}&
\sqrt{(1-\beta)d(1+\xi)\over(1-\beta+\xi)(1+d)}\cr
 0& -\sqrt{d\over 1+d}& \sqrt{1\over 1+d }\cr}.\labels{vg1}\ee
Similar to the case $G=2$, since $\tilde n_1=\tn_3$, we should
identify
$V_{31}=U_{e3}=s_{13}=0$, and hence $n_3=\nu_e$.
The atmospheric condition $\tan^2\theta_{13}=1$ then requires
$\xi=(1+d-\beta-d\beta)/(-1-d+2\beta+d\beta)$, and turns $\b V$ into
\be
\b V=\pmatrix{
\sqrt{1/(2+d)}& \sqrt{1/(2+d)}& \sqrt{d/(2+d)}\cr
-\sqrt{(1+d)/(2+d)}& \sqrt{1/(1+d)(2+d)}& \sqrt{d/(1+d)(2+d)}\cr
 0& -\sqrt{d/(1+d)}& \sqrt{1/(1+d)}\cr}.\labels{vg1a}\ee
Multiplying the second row and the second column each by $-1$, and 
interchanging the first and third rows and well as the first and third
columns, bring $\b V$ to the form
\be
\b U=\pmatrix{\sqrt{1/(1+d)}& \sqrt{d/(1+d)}&0\cr
 -\sqrt{d/(1+d)(2+d)}& \sqrt{1/(1+d)(2+d)}& \sqrt{(1+d)/(2+d)}\cr
\sqrt{d/(2+d)}&-\sqrt{1/(2+d)}&\sqrt{1/(2+d)}\cr},\labels{vg1b}\ee
which we can identify with $U$ of \eq{anglesg2} to fix the parameter
$d$ from the solar angle.

we can carry out a numerical fit like Table I and obtain
similar results. Since in this case $\rho\sim 1$ 
is a fairly good approximation
even for the $M_1=2$ situation of LMA, we will not show these fits here.

\section{Conclusion}
In theories with extra dimensions where ordinary particles 
are confined to the three-brane, only gravity and sterile
neutrino are allowed to roam in the bulk. Moreover, 
if there is only a single large extra dimension, 
the energy scale of gravity would still be too high 
to produce many gravitons,
then it leaves neutrino to be the only probe of extra dimensions
in high-energy experiments. It is therefore important to investigate
whether there are signs pointing to a higher dimension in the neutrino data.

A large extra dimension may explain the
small neutrino mass. Strong coupling between brane and bulk
neutrinos may also explain large mixing
not seen in the quark sector. Unfortunately, it is not easy to realize
these ideas, because oscillation into sterile neutrinos are greatly disfavored
in the solar and atmospheric neutrino data, and the presence of 
the Kaluza-Klein tower of sterile neutrinos from the bulk tends
to produce a very complicated oscillation pattern not hitherto observed.
In this paper we consider a simple and economic model where both of these
problems are solved. The model consists of a sterile
neutrino $\nu_s$ in additional to three active neutrinos $\nu_e,\nu_\mu,
\nu_\t$ on the brane, coupled strongly to a single common neutrino in
the bulk, but not directly among themselves. In the strong
coupling limit, the complicated oscillation patterns are washed out,
the induced couplings between the brane neutrinos become large,
 and the sterile
neutrinos on the brane and in the bulk can be made to decouple from the
active neutrinos. This results in a realistic theory of neutrino oscillations,
capable of accommodating all existing neutrino data. It also predicts
a natural connection between the smallness of the reactor angle
$\theta_{13}$, and the smallness of the solar to atmospheric mass gap
ratio.

This research is supported by the Natural Science and Engineering
Research Council of Canada and FCAR of Qu\'ebec.

\newpage

\begin{table}
\caption{Each of these tables shows the fitted
parameters $(\alpha,\beta,\gamma,\delta)$ for a given $M_1$ and $\xi$, when
$G=2$.  The resulting value of $\tan^2\theta_{13}$ is given.
The last two columns should be
equal to $\tan^2\theta_{12}$ and $\xi$ respectively 
in the $\rho=0$ limit. To save space, $1.8\x10^{-5}$, for
example, is written as 1.8(-5).}

\vskip1cm
LMA. $M_1=2$ eV. $\rho=.02$. $\tan^2\theta_{12}=0.36$.
\medskip

\begin{tabular}{cccccccc}
$\xi$&$\tan^2\theta_{13}$&
$\alpha$&$\beta$&$\gamma$&$\delta$&$\alpha/\beta$&$\gamma/(1-2\gamma)$\\ 
\hline 
.500&.0007&.0054&.0146&.2465&.7335&.3668&.4862\cr
.300&.0013&.0054&.0146&.1868&.7932&.3662&.2983\cr
.050&.0211&.0052&.0148&.0520&.9280&.3488&.0581\cr
.032&.0398&.0050&.0150&.0375&.9425&.3329&.0406\cr
\end{tabular}
\vskip1cm
LMA. $M_1=0$ eV. $\rho=.14$. $\tan^2\theta_{12}=0.36$.
\medskip

\begin{tabular}{cccccccc}
$\xi$&$\tan^2\theta_{13}$&
$\alpha$&$\beta$&$\gamma$&$\delta$&$\alpha/\beta$&$\gamma/(1-2\gamma)$\\ 
\hline
100&.0043&.0415&.0999&.4106&.4480&.4150&2.297\cr
 10&.0050&.0414&.1000&.3937&.4648&.4142&1.853\cr
  1&.0139&.0408&.1006&.2846&.5740&.4054&0.661\cr
0.34&.0391&.0390&.1024&.1901&.6685&.3807&0.307\cr
\end{tabular}
\vskip1cm
LOW. $M_1=2$ eV. $\rho=.00005$. $\tan^2\theta_{12}=0.69$.
\medskip

\begin{tabular}{cccccccc}
$\xi$&$\tan^2\theta_{13}$&
$\alpha$&$\beta$&$\gamma$&$\delta$&$\alpha/\beta$&$\gamma/(1-2\gamma)$\\ 
\hline
.5000&4.4(-9)&1.8(-5)&2.7(-5)&.2500&.7500&.6900&.4999\cr
.0100&4.9(-6)&1.8(-5)&2.7(-5)&.0098&.9901&.6900&.0100\cr
.0010&.0005&1.8(-5)&2.7(-5)&.0010&.9989&.6898&.0010\cr
.0001&.0313&1.8(-5)&2.7(-5)&.0001&.9998&.6738&.0001\cr
\end{tabular}
\vskip1cm

LOW. $M_1=0$ eV. $\rho=.00663$. $\tan^2\theta_{12}=0.69$.
\medskip

\begin{tabular}{cccccccc}
$\xi$&$\tan^2\theta_{13}$&
$\alpha$&$\beta$&$\gamma$&$\delta$&$\alpha/\beta$&$\gamma/(1-2\gamma)$\\ 
\hline
1&4.2(-5)&.0027&.0039&.3302&.6632&.6946&.9723\cr
.100&.0012&.0027&.0039&.0833&.9101&.6940&.0999\cr
.030&.0099&.0027&.0039&.0291&.9643&.6894&.0309\cr
.013&.0398&.0027&.0040&.0137&.9797&.6739&.0141\cr
\end{tabular}

\vskip1cm
VAC. $M_1=0$ eV. $\rho=.00024$. $\tan^2\theta_{12}=0.363$.
\medskip

\begin{tabular}{cccccccc}
$\xi$&$\tan^2\theta_{13}$&
$\alpha$&$\beta$&$\gamma$&$\delta$&$\alpha/\beta$&$\gamma/(1-2\gamma)$\\ 
\hline
.1000&1.3(-6)&6.3(-5)&.0002&.0834&.9164&.3631&.1001\cr
.0100&.0001  &6.3(-5)&.0002&.0099&.9899&.3630&.0101\cr
.0010&.0080  &6.2(-5)&.0002&.0011&.9987&.3562&.0011\cr
.0004&.0400  &5.9(-5)&.0002&.0005&.9993&.3292&.0005\cr
\end{tabular}
\vskip1cm

\end{table}

\newpage
\begin{figure}[h]
\includegraphics{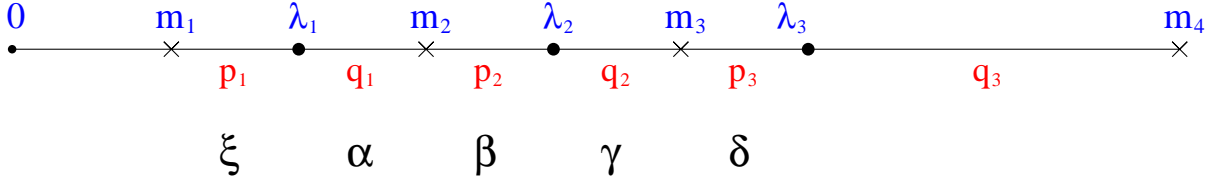}
\vspace{5cm}
\caption{A summary of the parameters in the strong-coupling
limit for $f=4$. $m_i$ are the
Majorana masses and $\l_\alpha$ are the isolated eigenvalues.
These parameters must be ordered as indicated.
Shown in the figures are also the distances $p_\alpha,q_\alpha$
as well as the ratios $\xi,\alpha,\beta,\gamma,$ and $\delta$.
The common distance with respect to which the ratios are taken is
$\l_3-\l_1$.}
\end{figure}

\end{document}